\documentclass[
    10pt
    ,aps
    ,pra
    ,oneside
    ,nobibnotes
    ,floatfix
    ,twocolumn
    ,nofootinbib
    ,superscriptaddress
    ,showkeys
]{revtex4-2}

\usepackage{amsmath,amssymb,bm}
\usepackage{array}
\allowdisplaybreaks[3]
\usepackage{siunitx}
\usepackage{graphicx}\graphicspath{{}{Graph/}{Graph/seminar/}
    {./Graph/}{./Graph/A0/}{./Graph/A1/}{./Graph/A2/}{./Graph/A3/}
    {../Graph/GlobalMode2/}
}

\usepackage[svgnames]{xcolor}

\usepackage{iftex}
\iftutex
    \usepackage{fontspec}

    \usepackage[math-style=ISO,bold-style=ISO]{unicode-math}
\else

    \usepackage[T1]{fontenc} 
    \usepackage[utf8]{inputenc} 
\fi

\usepackage[english, showlanguages]{babel}

    \iftutex
        
    \else
        
    \fi

    \DeclareMathOperator{\const}{const}

    \newcommand{\dif}[2][]{\mathop{}\!\mathrm{d}
        \if
            \relax\detokenize{#1}\relax
        \else
            ^{\mkern-1.mu#1}\mkern-2.5mu 
    \fi
    #2\,}
    \newcommand{\der}[2]{\frac{\dif{#1}}{\dif{#2}}}
    \newcommand{\tder}[2]{{\dif{#1}}/{\dif{#2}}}

    \newcommand{\mean}[1]{\left\langle #1\right\rangle}

\usepackage[
    colorlinks
        ,linkcolor=violet
        ,filecolor=purple
        ,citecolor=teal
        ,urlcolor=magenta
    ,unicode
    ,pdfpagemode=UseOutlines
    ,pdfdisplaydoctitle=true
    ,pdfpagetransition=Wipe
    ,pdfusetitle
]{hyperref}
\hypersetup{
        pdfkeywords={plasma, MHD stability, ballooning modes, mirror trap, Gas-Dynamic Trap, Compact Axisymmetric Toroid}
}
\usepackage{bookmark}
\usepackage[justification=centerlast]{caption}

\begin{document}

\def\bot{\mathrel\perp}



\title{
    Influence of the shape of a conducting chamber on the stability
    \\
    of rigid ballooning modes in a mirror trap
}

\author{Igor Kotelnikov}
    \email{I.A.Kotelnikov@inp.nsk.su}
    \affiliation{Budker Institute of Nuclear Physics, Novosibirsk, Russia}
\date{\today}

\begin{abstract}
    %
    The MHD stabilization of ``rigid'' flute and ballooning modes with azimuthal number $m = 1$ in an axisymmetric mirror trap by means of a perfectly conducting lateral wall is studied both in the presence and in the absence of the end MHD anchors. Numerical calculations were carried out for an anisotropic plasma created by injection of a beam of neutral atoms into the minimum of the magnetic field at the right angle to the trap axis. The stabilizing effect of the conducting shell in the form of a straight cylinder is compared with a proportional chamber, which, on an enlarged scale, repeats the shape of the plasma column.
    
    It is confirmed that for effective wall stabilization of the rigid modes, the plasma beta ($\beta$, the ratio of the plasma pressure to the magnetic field pressure) must exceed some critical value $\beta_{\text{crit}}$. When conducting lateral wall is combined with conducting end plates imitating MHD end anchors, there are two critical beta values and, respectively, two stability zones $\beta<\beta_{\text{ crit}1}$ and $\beta > \beta_{\text {crit}2}$ that can merge, making the entire range of allowable beta values $0<\beta<1$ stable.

    The dependence of the critical betas on the plasma anisotropy, mirror ratio, and the width of the vacuum gap between the plasma and the lateral wall is examined. In contrast to the earlier works of other authors focused on to the stepwise plasma model, the stability margins are calculated for a number of diffuse radial pressure profiles with different peakedness and several axial magnetic field profiles.

\end{abstract}


\maketitle

\section{Introduction}\label{s1}


Continuing the study of large-scale MHD instabilities, begun in our papers \cite{Kotelnikov+2022NF_62_096025,  Kotelnikov+2023NF_63_066027}, below we present new results of studying the so-called wall stabilization of rigid flute and ballooning perturbations with azimuthal number $m=1$ in a axially-symmetric mirror trap (also called linear or open trap) with anisotropic plasma. According to the physics of the case, wall stabilization of a plasma with a sufficiently high pressure is achieved by inducing image (Foucault's) currents in the conducting walls of the vacuum chamber surrounding the plasma column. These currents are directed oppositely to the diamagnetic currents in the plasma edge, and opposite currents are known to repel each other. Such repulsion returns the floating ``tongues'' of plasma back to the axis of the trap.

Under the conditions of a real experiment, small-scale flute and balloon oscillations are readily stabilized by finite Larmor radius effects (FLR effects) \cite{Rosenbluth+1962NFSuppl_1_143}. However, the FLR effects cannot stabilize oscillations with an azimuthal number $m=1$,but make them ``rigid'' in a certain sense. In two papers \cite{Kotelnikov+2022NF_62_096025, Kotelnikov+2023NF_63_066027} we successively analyzed the wall stabilization of isotropic and anisotropic plasmas in a model where the shape of a perfectly conducting lateral (side) wall on an enlarged scale repeated the shape of the plasma column, i.e.\ the ratio of the radius of the conducting wall $r_ {w}$ to the plasma radius $a$ is the same in all sections, $r_{w}/a=\const$. It is very difficult to manufacture a conducting chamber of this shape. Nevertheless, it is this model that has been used in most of the previous theoretical studies, since it simplifies the calculation. In the following, a conducting lateral wall of this shape will be called \emph{proportional chamber} and will be denoted by the abbreviation ``Pr''.


In this paper, we examine the stability of the flute-ballooning mode $m=1$ in a \emph{straight chamber} (denoted by the abbreviation ``St'') with a constant radius $r_{w} = \const$. We present the results of calculating the stability zones both in the absence of additional end MHD anchors, which are traditionally used to suppress flute oscillations in open axially symmetric traps, and in their presence.

In the first case, hereinafter referred to as \emph{lateral wall stabilization} and denoted by ``LW'' for brevity, there is one stability zone at a sufficiently large beta that exceeds the critical value, $\beta > \beta_{\text{crit}}$. In the second case, there are two such zones: the first one is for small beta, $\beta < \beta_{\text{crit}1}$, the second one is for large beta, $\beta > \beta_{\text{crit}2}$. These two zones can merge. We will call this case \emph{combined wall stabilization} and denote it by the abbreviation ``CW''.


Our goal is to calculate the critical values of $\beta_{\text{crit}}$, $\beta_{\text{crit}1}$, $\beta_{\text{crit}2}$ and identify the dependence of these values on the degree of plasma anisotropy, radial profile of plasma pressure, axial profile of the magnetic field, and the width of the vacuum gap between the plasma and the lateral wall. We also simulate MHD anchors by conducting end plates placed at different locations behind the throat of magnetic field plugs.

The same goals we had when examined earlier the case of proportional chamber. To simplify the comparison of stability zones for the proportional and straight chambers, we further use the same four radial pressure profiles and the same three magnetic field profiles as in Ref.~\cite{Kotelnikov+2022NF_62_096025, Kotelnikov+2023NF_63_066027}.

In an effort to reduce the inevitable duplication of part of the content of Ref.~\cite{Kotelnikov+2022NF_62_096025, Kotelnikov+2023NF_63_066027}, we skip the review of publications by other authors and restrict ourselves to a couple of remarks. First, we point out that the key equation that describes flute and balloon oscillations $m = 1$ in an open axially symmetric trap belongs to Lynda LoDestro \cite{LoDestro1986PF_29_2329}. Secondly, we  note that the effect of the shape of a conducting wall on its stabilizing properties was previously discussed by Li, Kesner and Lane \cite{LiKesnerLane1987NF_27_101}, and also by Li, Kesner and LoDestro \cite{LiKesnerLoDestro1987NF_27_1259}. The first work assumed that stabilizing wall extends axially only over a part of the distance between the mirror trap midplane and the throat. In a model of this arrangement, a wall was used which was near the plasma surface in the bad curvature region and distant from the plasma surface in the good curvature region. A variational method was used to solve the equations of a pre-LoDestro type for both regions assuming sharp-boundary radial profile of the plasma pressure. The authors of the second paper assumed that a certain length of magnetic field has a series of ripples in it. They concluded that with isotropic pressure wall stabilization is possible if the plasma beta value is higher than 50\%, provided that the conducting wall is very close to the plasma surface.



In section \ref{s2}, we reproduce the LoDestro equation with all the notation, but without detailed explanations, and from the very beginning we use dimensionless variables. Also, in dimensionless notation, in section \ref{s3} we repeat (with some changes) the formulas that model the anisotropic pressure distribution in a mirror trap under normal injection of neutral beams; the changes are due to the transition to a more economical parameterization in numerical calculations of the coefficients of the LoDestro equation. In sections \ref{s5} and \ref{s6}, the results of calculations are given first for the case when there are no other means of MHD stabilization besides the lateral conducting wall, and then for the case when conducting plates are installed in magnetic mirrors or behind them, which simulate the effect of end MHD anchors of different strength. Finally, section \ref{s9} summarizes our observations and conclusions.

\section{LoDestro equation}\label{s2}

The LoDestro equation is a second-order ordinary differential equation for the function
    \begin{equation}
    \label{2:02}
    \phi(z) = a(z) B_{v}(z) \xi_{n}(z)
    ,
    \end{equation}
which depends on single coordinate $z$ along the trap axis and is expressed in terms of the variable radius of the plasma column boundary $a=a(z)$, the vacuum magnetic field $B_{v}=B_{v}(z)$ and the virtual small displacement $ \xi_{n}=\xi_{n}(z)$ of the plasma column from the axis.
%
In its final form, the LoDestro equation reads
    \begin{multline}
    \label{2:01}
    0 = \der{}{z}
    \left[
        \Lambda + 1 - \frac{2\mean{\overline{p}}}{B_{v}^{2}}
    \right]
    \der{\phi}{z}
    \\
    +
    \phi
    \left[
        - \der{}{z}\left(
            \frac{B_{v}'}{B_{v}} + \frac{2a'}{a}
        \right)
        \left(
            1 - \frac{\mean{\overline{p}}}{B_{v}^{2}}
        \right)
    +
    \frac{\omega^{2}\mean{\rho}}{B_{v}^{2}}
    \right.
    \\
    \left.
    -
    \frac{2\mean{\overline{p}}}{B_{v}^{2}}\frac{a_{v}''}{a_{v}}
    -
    \frac{1}{2}\left(
            \frac{B_{v}'}{B_{v}} + \frac{2a'}{a}
    \right)^{2}
    \left(
        1 - \frac{\mean{\overline{p}}}{B_{v}^{2}}
    \right)
    \right]
    ,
    \end{multline}
where the derivative $\tder{}{z}$ in the first two lines acts on all factors to the right of it, the prime ($'$) is a shortcut for $\tder{}{z}$, and $\omega $ is the oscillation frequency.
%
Other notations are defined as follows
    \begin{gather}
    \label{2:03}
    \frac{a^{2}}{2} = \int_{0}^{1} \frac{\dif{\psi}}{B}
    ,\\
    \label{2:03a}
    \frac{r^{2}}{2} = \int_{0}^{\psi} \frac{\dif{\psi}}{B}
    ,\\
    \label{2:04}
        B^{2} = B_{v}^{2} -2p_{\bot}
    ,\\
    \label{2:07}
    a_{v}(z) = \sqrt{\frac{2}{B_{v}(z)}}
    ,\\
    \label{2:05}
    \overline{p} = \frac{p_{\bot} + p_{\|}}{2}
    ,\\
    \label{2:06}
    \mean{\overline{p}}
    =
    \frac{2}{a^{2}}
    \int_{0}^{1} \frac{\dif{\psi}}{B}\,\overline{p}
    ,\\
    \label{2:09}
    \Lambda = \frac
    {
        r_{w}^{2} + a^{2}
    }{
        r_{w}^{2} - a^{2}
    }
    .
    \end{gather}
Equation \eqref{2:03a} relates the radial coordinate $r$ and the magnetic flux $\psi$ through a ring of radius $r$ in the $z$ plane. The magnetic field $B=B(\psi,z)$, weakened by the plasma diamagnetism, in the paraxial (long-thin) approximation (i.e., with a small curvature of field lines) is related to the vacuum magnetic field $B_{v}=B_ {v }(z)$ by the transverse equilibrium equation \eqref{2:04}.
Kinetic theory predicts (see, for example, \cite{Newcomb1981JPP_26_529}) that the transverse and longitudinal plasma pressures can be considered as functions of magnetic field $B$ and magnetic flux $\psi$, i.e.\ $p_{\bot}=p_{\bot}(B ,\psi)$, $p_{\|}=p_{\|}(B,\psi)$. In Eq.~\eqref{2:01}, one must assume that the magnetic field $B$ is already expressed in terms of $\psi$ and $z$, and therefore $p_{\bot}=p_{\bot}(\psi,z)$, $p_{\|}=p_{\|}(\psi,z)$. The angle brackets in Eq.~\eqref{2:01} denote the mean value of an arbitrary function of $\psi$ and $z$ over the plasma cross section. In particular, the average value $\mean{\rho}$ of the density $\rho=\rho(\psi,z)$ is calculated using a formula similar to \eqref{2:06}.


Parameter $\Lambda $ is, generally speaking, a  function of the $z$ coordinate. It implicitly depends on the plasma parameters and the magnetic field through the dependence of the plasma column radius $a=a(z)$ on them. In the special case of a proportional chamber, when $r_{w}(z)/a(z)=\const$, the function $\Lambda(z)$ becomes a constant, which simplifies the equation somewhat. Namely, for the sake of such simplification, in our previous studies we assumed that $\Lambda =\const$. Now we will abandon this simplification and consider a completely realistic case of a conducting chamber in the form of a straight cylinder, that is, we will assume that $r_{w}=\const$. Comparison of the sizes of the stability zones for the proportional and straight conducting chambers, which in a sense are antipodes, will allow us to estimate how strong the influence of the shape of the conducting chamber on the stability of ballooning oscillations is.

Traditionally, two types of boundary conditions are considered. In the presence of conducting end plates straightly in the magnetic mirrors at $z\pm1$, the boundary condition
     \begin{equation}
     \label{2:11}
     \phi(\pm1) = 0
     \end{equation}
should be chosen. In a more general case, when the conducting end plate is installed somewhere in the behind-the-mirror region, namely, in the plane with coordinates $z=\pm z_{e}$, the zero boundary condition must obviously be assigned to this plane:
     \begin{equation}
     \label{2:11S}
     \phi(\pm z_{e}) = 0
     .
     \end{equation}
By solving the LoDestro equation with the boundary condition \eqref{2:11S}, it is possible to simulate the effect of end MHD anchors with different stability margins.

If the plasma ends are electrically isolated, the boundary condition
    \begin{equation}
    \label{2:12}
    \phi'(\pm1) = 0
    \end{equation}
is applied at $z=\pm1$. As a rule, it implies that other methods of MHD stabilization in addition to stabilization by a conducting lateral wall are not used.


An obvious fit to the LoDestro equation with boundary conditions \eqref{2:11}, \eqref{2:11S} or \eqref{2:12} is the trivial solution $\phi\equiv0$. To eliminate the trivial solution, we impose a normalization in the form of one more condition
    \begin{equation}
    \label{5:03}
    \phi(0)=1.
    \end{equation}
%
Taking into account the symmetry of the magnetic field in actually existing open traps with respect to the median plane $z=0$, it suffices to find a solution to the LoDestro equation at a half the distance between the magnetic mirrors, for example, in the interval $0<z<1$. Due to the same symmetry, the desired function $\phi(z)$ must be even, therefore
    \begin{equation}
    \label{5:01}
    \phi'(0)=0.
    \end{equation}
%
It is convenient to search for a solution to the LoDestro equation by choosing the boundary conditions \eqref{5:03} and \eqref{5:01}. In theory, a second-order linear ordinary differential condition with two boundary conditions must always have a solution. However, the third boundary condition \eqref{2:11}, \eqref{2:11S} or \eqref{2:12} can only be satisfied for a certain combination of the problem parameters. If the parameters of the plasma, magnetic field, and geometry of the lateral conducting wall are given, the third boundary condition should be considered as a nonlinear equation for the frequency squared $\omega^{2}$. If the root of such an equation is greater than zero, MHD oscillations with azimuthal number $m=1$ are stable; if $\omega ^{2}<0$, then instability takes place. On the margin of the stability zone $\omega ^{2}=0$. In this case, the solution of the boundary-value problem \eqref{2:01}, \eqref{5:03}, \eqref{5:01} with the additional boundary condition \eqref{2:12} or \eqref{2:11} gives the critical value of beta, respectively $\beta_{\text{crit}}$ or $\beta_{\text{crit}1}$ and $\beta_{\text{crit}2}$.

As it was mentioned in Ref.~\cite{Kotelnikov+2023NF_63_066027, Kotelnikov+2023NF_63_066027}, the LoDestro equation \eqref{2:01} with boundary conditions \eqref{2:11}, \eqref{2:11S}  or \eqref{2:12} constitutes the standard Sturm-Liouville problem. At first glance, it may seem that the solution of such a problem is rather standard. However, the equation has the peculiarity that its coefficients can be singular. In the anisotropic pressure model, formulated in Section \ref{s3}, the singularity appears near the minimum of the magnetic field in the limit $\beta \to 1$. By some indications, it can be assumed that our predecessors were aware of the singularity problem. Unfortunately, they did not leave us recipes for dealing with this singularity.

\section{Plasma pressure and magnetic field}\label{s3}

In this article, we study wall stabilization using the same model of an anisotropic plasma that was introduced in our previous article \cite{Kotelnikov+2022NF_62_096025, Kotelnikov+2023NF_63_066027}. It approximately simulates the pressure distribution in an open trap with normal (transverse) injection of neutral beams (NB). Its undoubted advantage is that it allows a significant part of the calculations to be performed in an analytical form. In addition, such a pressure distribution is not subject to mirror and firehose instabilities. Below we repeat the necessary formulas, writing them in dimensionless variables, such that $B=B_{v}=1$ at the ``stop'' point, where the pressure vanishes:
    \begin{gather}
    \label{3:08}
    p_{\bot}(B,\psi)
    =
    p_{0}f_{k}(\psi)
    \left(1-B^2\right)
    ,\\
    \label{3:09}
    p_{\|}(B,\psi)
    =
    p_{0}f_{k}(\psi)
    \left(1-B\right)^{2}
    ,\\
    \label{3:10}
    \overline{p}(B,\psi) =
    p_{0}f_{k}(\psi)
    \left(1-B\right)
    .
    \end{gather}
It is assumed here and below that the entire region between the stop point and magnetic mirror throat is occupied by relatively cold plasma, so that $p_{\bot}=p_{\|}=\overline{p}=0$ for $B>1$. Dimensionless function $f_{k}(\psi)$ describes radial profile of the plasma pressure. It is defined as
    \begin{equation}
    \label{3:22}
    f_{k}(\psi) =
    \begin{cases}
      1 - \psi^{k}, & \mbox{if } 0\leq \psi \leq 1 \\
      0, & \mbox{otherwise}
    \end{cases}
    \end{equation}
for integer values of index $k$, and for $k=\infty$
is expressed in terms of a $\theta$-function such that $\theta(x)=0$ for $x<0$ and $\theta(x)=1$ for $x>0$:
    \begin{equation}
    \label{3:23}
    f_{\infty}(\psi) = \theta(1-\psi)
    .
    \end{equation}
For the model \eqref{3:08}, Eq.~\eqref{2:04} can be solved to explicitly express the magnetic field $B$ in terms of the vacuum field $B_{v}$:
    \begin{gather}
    \label{3:11}
    B(\psi,z)=
    \sqrt{
        \frac{B_{v}^{2}(z)-2p_{0}f_{k}(\psi)}{1-2p_{0}f_{k}(\psi)}
    }
    .
    \end{gather}

%
Choosing the function $B_{v}(z)$ for the calculations presented below, we use the second of the two models that were use in our first work \cite{Kotelnikov+2022NF_62_096025}:
    \begin{equation}
    \label{4:35}
    B_{v}(z) = \left[
        1 + (M-1)\sin^{q}(\pi z/2)
    \right]/R
    .
    \end{equation}
%
Note that it differs from original model of Ref.~\cite{Kotelnikov+2022NF_62_096025} by the factor $1/R$, which is associated with a change in the method of nondimensionalizing the function $B_{v}(z)$. Earlier we normalized the vacuum magnetic field to its minimum value $ B_{v}(0)$ in the median plane of the trap, $z=0$. 
Later it was noticed that in order to save computer resources it is more convenient to normalize the magnetic field on its magnitude at the stop point. This change was first implemented in Ref.~\cite{Kotelnikov+2023NF_63_066027}. It has already been taken into account in the equations~\eqref{3:08}–\eqref{3:10}. With the new normalization we have $\min(B_{v})=B_{v}(0)=1/R$, $\max(B_{v})=B_{v}(\pm1)=M/R$.

\begin{figure}
  \centering
  \includegraphics[width=\linewidth]{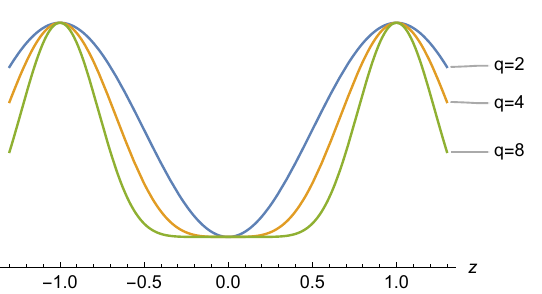}
  \caption{
    Axial profile of the vacuum magnetic field \eqref{4:35} for mirror ratio $K=8$ and three values of the index $q$ indicated on the graphs.
  }\label{fig:Bv_vs_z_q}
\end{figure}
\begin{figure}
  \centering
  \includegraphics[width=\linewidth]{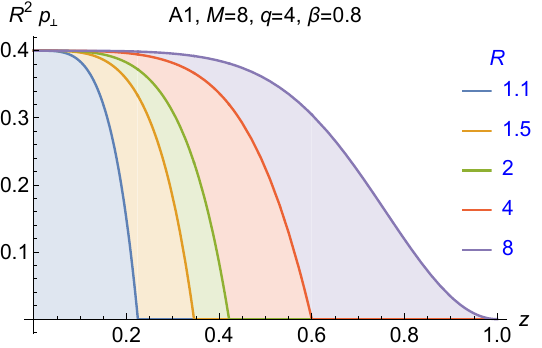}
  \includegraphics[width=\linewidth]{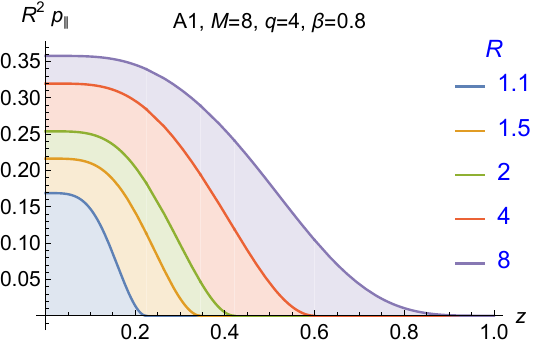}
  \caption{
  Anisotropic pressure model A1.
    Axial profiles of the transversal  \eqref{3:08}  and longitudinal pressure \eqref{3:09} in the magnetic field \eqref{4:35} for mirror ratio $M=8$, index $q=4$ and various mirror ratios $R$ at the stop points where pressure of hot plasma component drops to zero.
  }\label{fig:GM2-A1-Pt_vs_R-M8_q4}
\end{figure}

Parameter $R$ has the meaning of the mirror ratio between the stop point $B=B_{v}=1$, where the plasma pressure drops to zero, and the minimum value of the vacuum magnetic field in the middle plane of the trap. It determines the width of the pressure peak near the midplane as shown in Fig.~\ref{fig:GM2-A1-Pt_vs_R-M8_q4}. It can also serve as a measure of plasma anisotropy. The anisotropy is the greater, the smaller $R$, reaching a maximum at $R\to 1$.

Parameter $M=\max(B_{v})/\min(B_{v})$ is equal to the vacuum mirror ratio in the traditional sense.

Parameter $q$ in Eq.~\eqref{4:35} determines the steepness and width of the magnetic mirrors: the larger $q$, the smaller the fraction of the mirror area in the total length of the trap, as shown in as shown if Fig.~\ref{fig:Bv_vs_z_q}.

%
Parameter beta $\beta $ is defined below as the maximum of the ratio $2p_{\bot}/B_{v}^{2}$,
Parameter beta, $\beta $, is defined as the maximum of the ratio $2p_{\bot}/B_{v}^{2}$,
    \begin{gather}
    \beta=\max(2p_{\bot}/B_{v}^{2})
    .
    \end{gather}
The maximum is reached on the trap axis (where $\psi=0$) at the vacuum field minimum (where $\min(B_{v})=1/R$), so that
    \begin{gather}
    \label{3:18}
    \beta
    =
    \frac{2 \left(R^2-1\right)p_{0}}{1 - 2p_{0}}
    .
    \end{gather}
%
Parameter $p_{0}$ can vary within $0<p_{0}<1/2R^{2}$, and $\beta \to 1$ for $p_{0}\to 1/2R^{2} $. For $\beta >1$, plasma equilibrium is impossible, since Eq.~\eqref{2:04} does not have a continuous solution.

In our internal classification, the isotropic plasma variant is designated `A0', the anisotropic pressure \eqref{3:08}–\eqref{3:10}, which is formed during normal neutral beam injection (NBI), is designated `A1'.  Variations with conducting lateral wall stabilization will be marked with the letters `LW' (shortcut for Lateral Wall), and variations with combined lateral wall stabilization and conducting end plates installed in the throat of the magnetic plug will be marked with the letters `CW' (shortcut for  Combined Wall). The design of a proportional chamber will be labeled `Pr' (Proportional). Thus, the label `A1-LWPr' in a figure caption corresponds to the variant of plasma stabilization with anisotropic pressure of the `A1' type in a proportional conducting chamber without the effect of the end MHD stabilizers. The abbreviation `St' (Straight) is assigned to a straight chamber.

\section{Stabilization by lateral wall}\label{s5}

Equation \eqref{2:01}, supplemented by the boundary conditions \eqref{5:03}, \eqref{5:01} and \eqref{2:12}, constitutes the Sturm-Liouville problem on half the interval $0\leq z \leq $1 between magnetic plugs. Its solution at $\omega^{2}=0$ gives the critical value of beta, $\beta_{\text{crit}}$, in the case when end MHD anchors are not used for MHD stabilization. The peculiarities of solving this problem for the models of anisotropic pressure and vacuum magnetic field adopted by us are described in detail in Ref.~\cite{Kotelnikov+2023NF_63_066027, Kotelnikov+2023NF_63_066027}, so we immediately proceed to the analysis of the results obtained.

Our calculations were carried out for the magnetic field \eqref{4:35} with mirror ratios chosen from the set $M\in\{24,16,8,4,2\}$ for most of possible commutations of the parameters $k\in\{1,2,4,\infty\}$, $q\in\{2,4,8\}$ and discrete values $\Lambda(0)=\{1,1.001,1.002,1.003, \ldots, 400,450,500\}$ of the function $\Lambda(z) $ at $z=0$. Parameter $R$ varied from $R=1.1$ to $R=M$ taking discrete values, as a rule, from the set $R\in\{1.1,1.2,1.5,2,4,8,16,24\}$.


\begin{figure*}
\includegraphics[width=\linewidth]{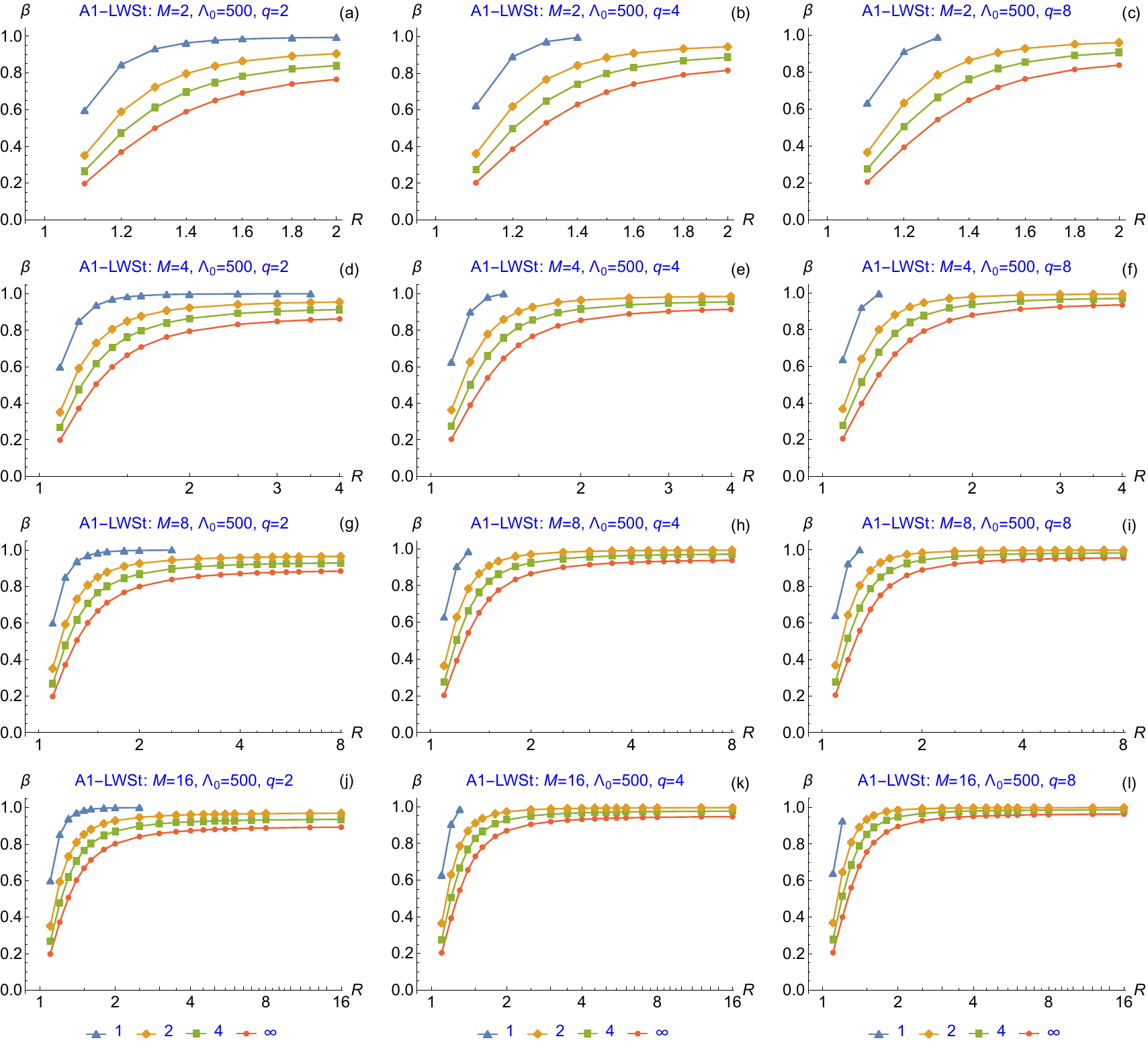}
  \caption{
    Critical beta in model magnetic field \eqref{4:35} versus the stop point mirror ratio $R$ for a set of mirror ratios $M$ and indices $q$ in the limit $\Lambda(0) \to \infty $. Stability zone for the radial profile with index $k$ is located above the corresponding curve.
    Compare with Fig.~2\textsuperscript{\cite{Kotelnikov+2023NF_63_066027, Kotelnikov+2023NF_63_066027}} in Ref.~\cite{Kotelnikov+2023NF_63_066027, Kotelnikov+2023NF_63_066027}.
  }
  \label{fig:GM2-A1-beta_vs_R-L500}
\end{figure*}

First of all, we compute the critical beta $\beta_{\text{crit}}$ in straight chamber at the closest location of the conducting wall to the surface of the plasma column to compare it with similar computation in Ref.~\cite{Kotelnikov+2023NF_63_066027} for proportional chamber. Figure \ref{fig:GM2-A1-beta_vs_R-L500} shows a series of graphs for the case $\Lambda(0)=500$, when the conducting wall almost touches the plasma column surface in its widest section $z=0$. They illustrate the dependence of the critical beta on parameter $R$, which characterizes the degree of plasma anisotropy (anisotropy is stronger for smaller $R$) and the spatial width of the pressure peak (the peak becomes wider for larger $R$). Within each figure, it is not difficult to detect a trend towards a decrease in the critical value of beta with an increase in the steepness (sharpness) of the radial pressure profile as the index $k$ increases from $k=1$ to $ k=\infty$ for a fixed pair of parameters $q$ and $M$. Comparison with a similar series of graphs in Fig.~2\textsuperscript{\cite{Kotelnikov+2023NF_63_066027}} in Ref.~\cite{Kotelnikov+2023NF_63_066027}\footnote{
    Numbers of figures in other articles hereinafter are marked with the reference number to the corresponding article in the bibliography.
} shows that the critical beta is definitely lower in the case of a proportional chamber over the entire range of the parameter $R$. This fact is especially noticeable for a sufficiently large value of $R$ and has a completely obvious explanation. It should be understood that graphs with the same labels (a), (b), (c), \ldots\ in the indicated figures correspond to the same combinations of parameters $q$ and $M$, but Fig.~2\textsuperscript{\cite{Kotelnikov+2023NF_63_066027, Kotelnikov+2023NF_63_066027}}  is built on the assumption that the vacuum gap between the conducting wall and the plasma has zero width throughout the plasma column, while Fig.~\ref{fig:GM2-A1-beta_vs_R-L500} in this article reflects the situation when the gap is minimal only in the median section of the plasma column, where its radius is maximum, but the gap gradually increases as the cross section coordinate approaches the magnetic mirrors, which of course weakens the image currents in the walls of the conducting chamber.


The second observation is that the straight chamber stabilizes the smoothest radial pressure profile with index $k=1$ noticeably worse, in the sense that the range of values of the $R$ parameter at which this profile can be stabilized is noticeably narrower. In the next section, we will see that in the presence of end MHD anchors, on the contrary, the smoothest radial profile is the easiest to stabilize.

\begin{figure*}
\includegraphics[width=\linewidth]{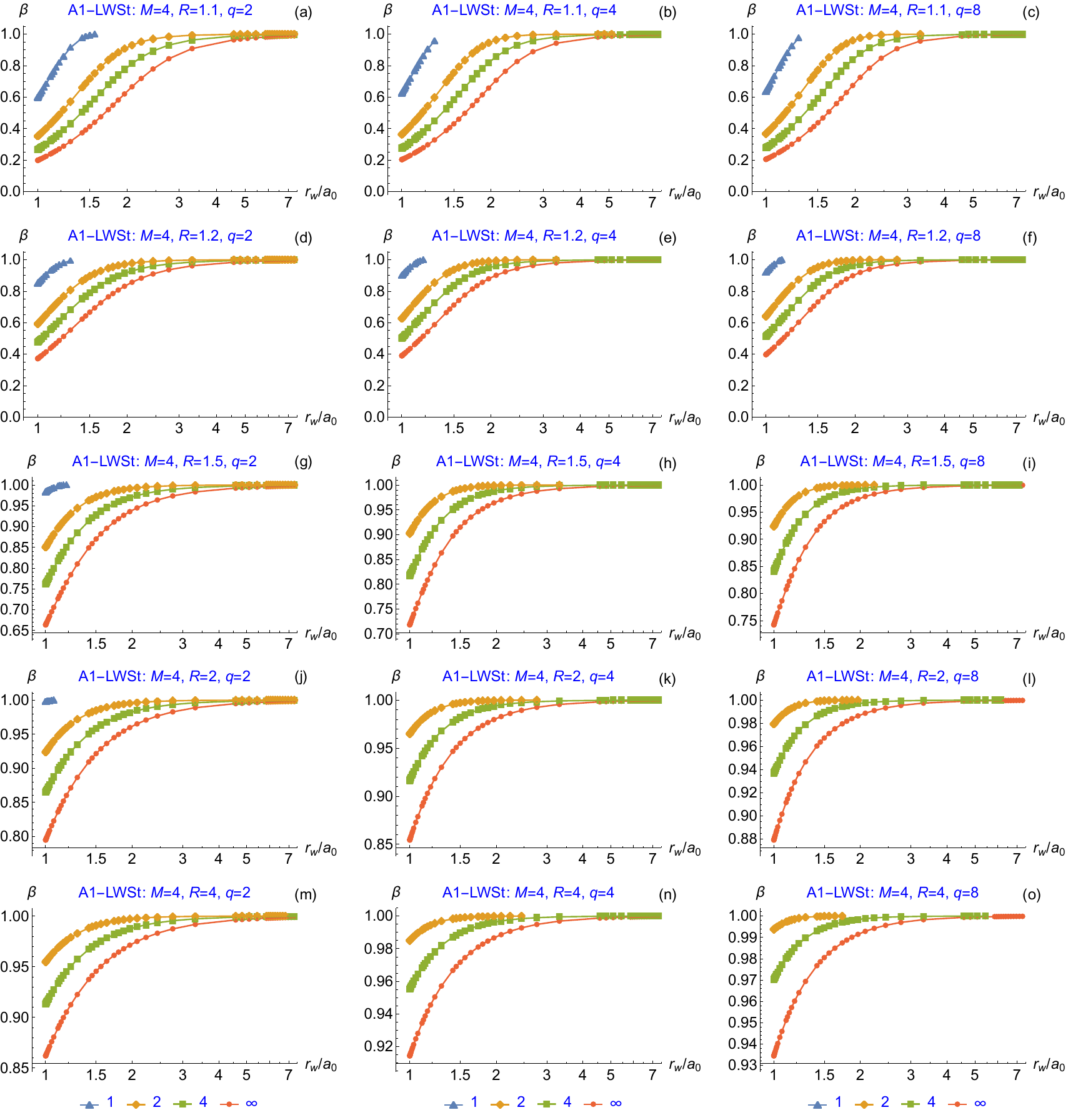}
  \caption{
    Critical beta for model magnetic field \eqref{4:35}, straight chamber and anisotropic plasma pressure model \eqref{3:08}, simulating normal NBI; $M = 4$, $R \in \{1.1, 1.2, 1.5, 2, 4\}$.
    Correspondence of the colors and markers of the curves to the index $k$ is shown under the bottom row of the graphs.
    Compare with Fig.~5\textsuperscript{\cite{Kotelnikov+2023NF_63_066027, Kotelnikov+2023NF_63_066027}} in Ref.~\cite{Kotelnikov+2023NF_63_066027, Kotelnikov+2023NF_63_066027}.
  }
  \label{fig:GM2-A1-LWSt_beta_vs_rw_setAll_M4qs}
\end{figure*}
\begin{figure*}
\includegraphics[width=\linewidth]{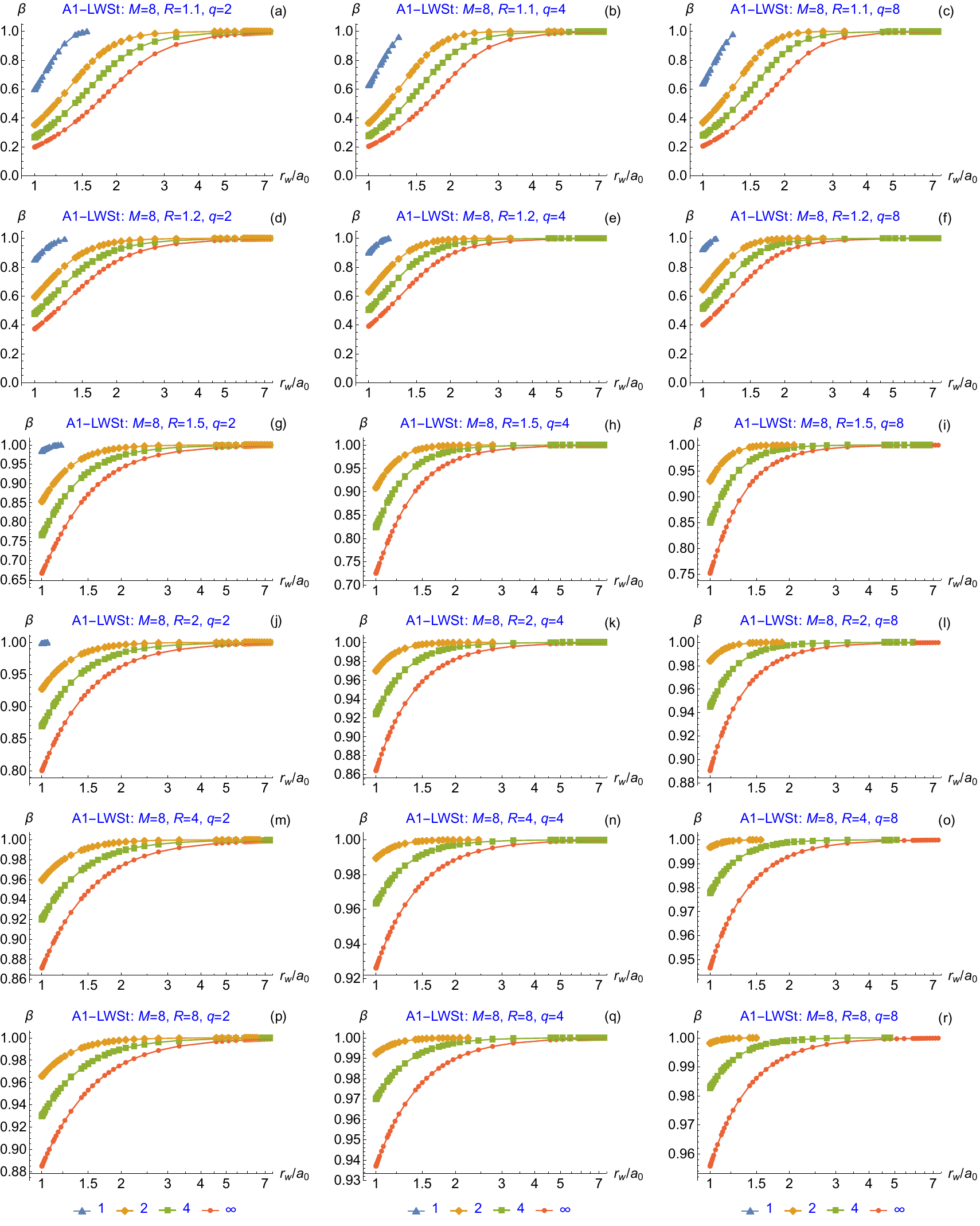}
  \caption{
    Critical beta for model magnetic field \eqref{4:35}, straight chamber and anisotropic plasma pressure model \eqref{3:08}, simulating normal NBI; $M = 8$, $R \in \{1.1, 1.2, 1.5, 2, 4, 8\}$.
    Correspondence of the colors and markers of the curves to the index $k$ is shown under the bottom row of the graphs.
    Compare with Fig.~6\textsuperscript{\cite{Kotelnikov+2023NF_63_066027, Kotelnikov+2023NF_63_066027}} in Ref.~\cite{Kotelnikov+2023NF_63_066027, Kotelnikov+2023NF_63_066027}.
  }\label{fig:GM2-A1-LWSt_beta_vs_rw_setAll_M8qs}
\end{figure*}
\begin{figure*}
\includegraphics[width=\linewidth]{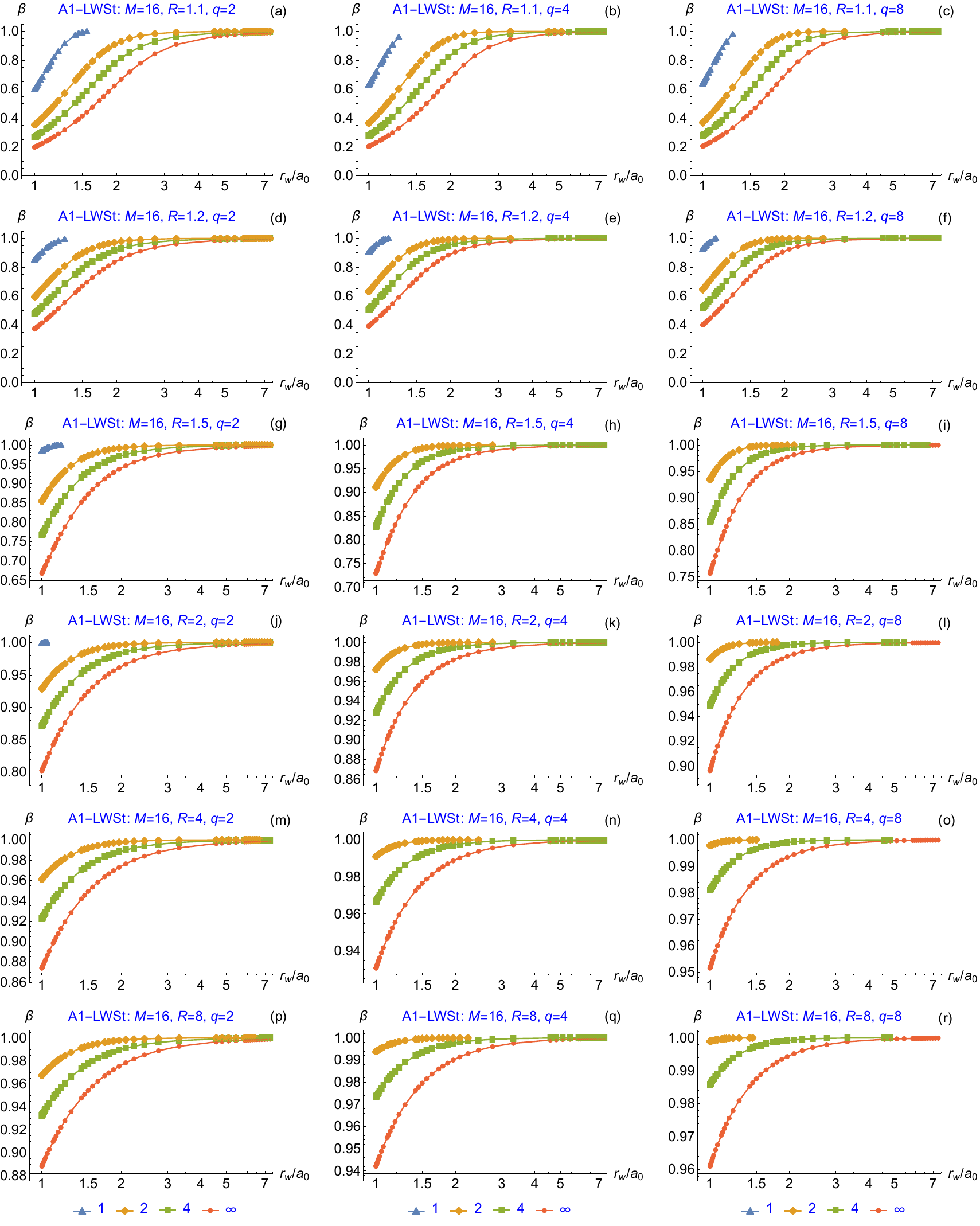}
  \caption{
    Critical beta for model magnetic field \eqref{4:35}, straight chamber and anisotropic plasma pressure model \eqref{3:08}, simulating normal NBI; $M = 16$, $R \in \{1.1, 1.2, 1.5, 2, 4, 8, 16\}$.
    Correspondence of the colors and markers of the curves to the index $k$ is shown under the bottom row of the graphs.
    Compare with Fig.~7\textsuperscript{\cite{Kotelnikov+2023NF_63_066027, Kotelnikov+2023NF_63_066027}} in Ref.~\cite{Kotelnikov+2023NF_63_066027, Kotelnikov+2023NF_63_066027}.
   }\label{fig:GM2-A1-LWSt_beta_vs_rw_setAll_M16qs}
\end{figure*}

Figures \ref{fig:GM2-A1-LWSt_beta_vs_rw_setAll_M4qs}, \ref{fig:GM2-A1-LWSt_beta_vs_rw_setAll_M8qs}, and \ref{fig:GM2-A1-LWSt_beta_vs_rw_setAll_M16qs} contain graphs of $\beta_{\text{crit}}$ versus  ratio $r_{w}/a_{0}=\sqrt{(\Lambda(0)+1)/(\Lambda(0)-1)}$ for mirror ratios $M\in\{4,8,16\}$ at various combinations of $q$ and $R$. Comparison of graphs in consecutive rows once more demonstrate strong dependence of critical betas on parameter $R$. Comparison of graphs within each row confirms a weak tendency noted in Ref.~\cite{Kotelnikov+2023NF_63_066027, Kotelnikov+2023NF_63_066027} to increase critical beta as the magnetic mirrors steepen with increasing parameter $q$. This effect is more evident for larger values of $R$, since they correspond to a more wide distribution of plasma pressure, which ``feels'' larger part of the device where vacuum gap in case of straight chamber is wider than in the midplane.

Comparison of Figs.~\ref{fig:GM2-A1-LWSt_beta_vs_rw_setAll_M4qs}–\ref{fig:GM2-A1-LWSt_beta_vs_rw_setAll_M16qs} between themselves shows a moderate decrease of the stability zone (located above the curves) as the mirror ratio $M$ grows. Again, this effect is more pronounced for larger $R$.


To evaluate the effect of the conducting wall shape on the wall stabilizing properties, one should compare figures \ref{fig:GM2-A1-LWSt_beta_vs_rw_setAll_M4qs}, \ref{fig:GM2-A1-LWSt_beta_vs_rw_setAll_M8qs} and \ref{fig:GM2-A1-LWSt_beta_vs_rw_setAll_M16qs} with figures 5\textsuperscript{\cite{Kotelnikov+2023NF_63_066027, Kotelnikov+2023NF_63_066027}}, 6\textsuperscript{\cite{Kotelnikov+2023NF_63_066027, Kotelnikov+2023NF_63_066027}}, 7\textsuperscript{\cite{Kotelnikov+2023NF_63_066027, Kotelnikov+2023NF_63_066027}} in Ref.~\cite{Kotelnikov+2023NF_63_066027, Kotelnikov+2023NF_63_066027}, respectively. The difference in the magnitude of $\beta_{\text{crit}}$ is hardly noticeable at $R=1.1$ and gradually increases as $R$ approaches $M$. For example, for $M=R=8$, $q=2$ and $r_{w}/a=r_{w}/a_{0}=1$ the value of $\beta_{\text{crit}}$ for all indices $k$ is larger by about $5\%$ for straight chamber than for proportional one. Such a difference does not seem critically large. Again, it should be clarified that if the values of the parameters $R$, $r_{w}/a$ in the figures in Ref.~\cite{Kotelnikov+2023NF_63_066027, Kotelnikov+2023NF_63_066027} are equal to the values of the parameters $M$, $R$ , $r_{w}/a_{0}$ in the figures in this article, the average (along the length of the plasma column) width of the vacuum gap in the case of a proportional chamber will be smaller than in the case of a straight chamber. That is why it seems that the straight chamber stabilizes the plasma worse, but this deterioration does not seem to be critically large. If we compare the average width of the vacuum gaps, then this deterioration will be less noticeable.

%

The minimum $\beta_{\text{crit}}$ for the studied set of parameters varies from $\num{0.197872}$ to $\num{0.197761}$ for mirror ratios in the range from $M=16$ to $ M=4$. It is obtained for $R=1.1$ and $q=2$, whereas in the case of a proportional chamber we had \num{0.178623} for $M=16$ and \num{0.178558} for $M=4$. This difference seems to be insignificant.


It can be assumed that under the conditions of a real experiment, the shape of the conducting chamber will be something in between the proportional and straight chambers. In this sense, these two chambers are two antipodes, and the results of calculations for these chambers determine the boundaries of the interval of the expected value $\beta_{\text{crit}}$ for a real experiment. However, it is currently not entirely clear whether abrupt changes in the radius $r_{w}(z)$ of the conducting chamber can have a strong stabilizing effect. Such changes can imitate the placement of a massive limiter near the plasma column in the region of unfavorable curvature of magnetic field lines.


Summing up this section, we can conclude that although MHD plasma stabilization by means of only perfectly conducting lateral wall without the use of accompanying stabilization methods is possible in principle, it requires either a sufficiently strong plasma anisotropy or sufficiently large values of beta. In this case, a natural question arises, how to transfer the plasma to such a state through the intermediate stages of a startup.




\clearpage

\section{Combined wall stabilization}\label{s6}



To adapt our calculation to the case of wall stabilization in combination with MHD end anchors, we should replace the boundary condition \eqref{2:12} with the boundary condition \eqref{2:11}, which means that the plasma is frozen into the conducting end plates. And again, as mentioned in section \ref{s5}, the features of solving this problem for the model of anisotropic pressure and vacuum magnetic field adopted in this paper are described in detail in Ref.~\cite{Kotelnikov+2023NF_63_066027, Kotelnikov+2023NF_63_066027}. There, formulas were obtained that refine the boundary conditions in a form suitable for a conducting chamber of any shape, so we will immediately proceed to the analysis of the results obtained.

\begin{figure*}
\includegraphics[width=\linewidth]{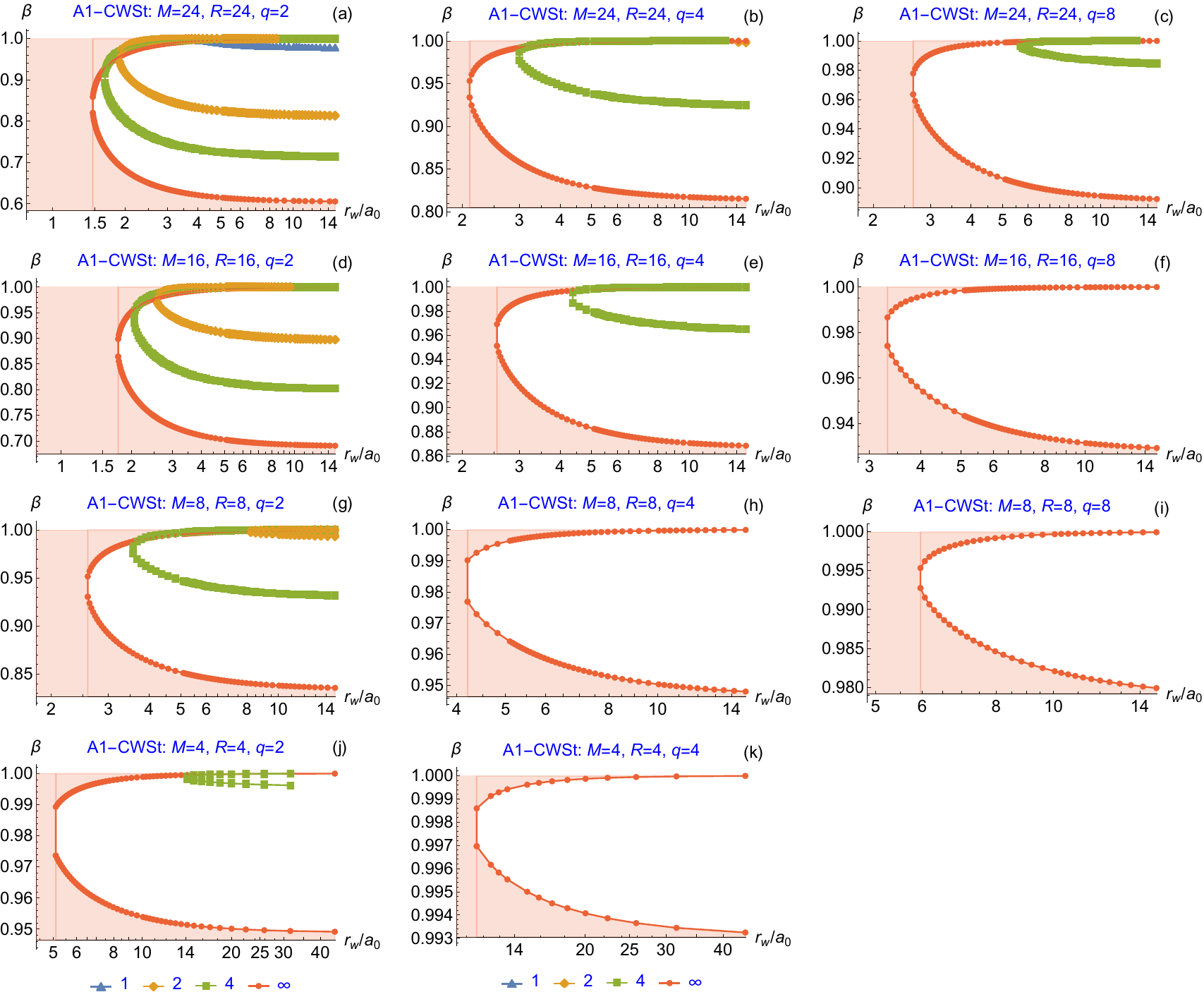}
  \caption{
    Stability map A1-CWSt for model magnetic field (36), anisotropic plasma pressure model (12) simulating normal NBI, straight conducting chamber combined with end MHD anchors;
    $q\in\{2, 4, 8\}$, $M\in\{24, 16, 8, 4\}$ and minimum anisotropy $R=M$. The unstable zone is located between the lower curve
    $\beta_{\text{crit}1}(r_{w}/a_{0})$ and the upper curve $\beta_{\text{crit}2}(r_{w}/a_{0})$ of one color; the stability zone is shaded for a plasma with a sharp boundary ($k = \infty $), for which it has the minimum dimensions; correspondence of the index $k$ to the color of curves is shown at the bottom of the figure.
    Compare with Figs.~9\textsuperscript{\cite{Kotelnikov+2023NF_63_066027, Kotelnikov+2023NF_63_066027}} and~10\textsuperscript{\cite{Kotelnikov+2023NF_63_066027, Kotelnikov+2023NF_63_066027}} in Ref.~\cite{Kotelnikov+2023NF_63_066027, Kotelnikov+2023NF_63_066027}.
  }
  \label{fig:GM2-A1-CWSt_beta_vs_rw_M=R}.
\end{figure*}

\begin{figure*}
\includegraphics[width=\linewidth]{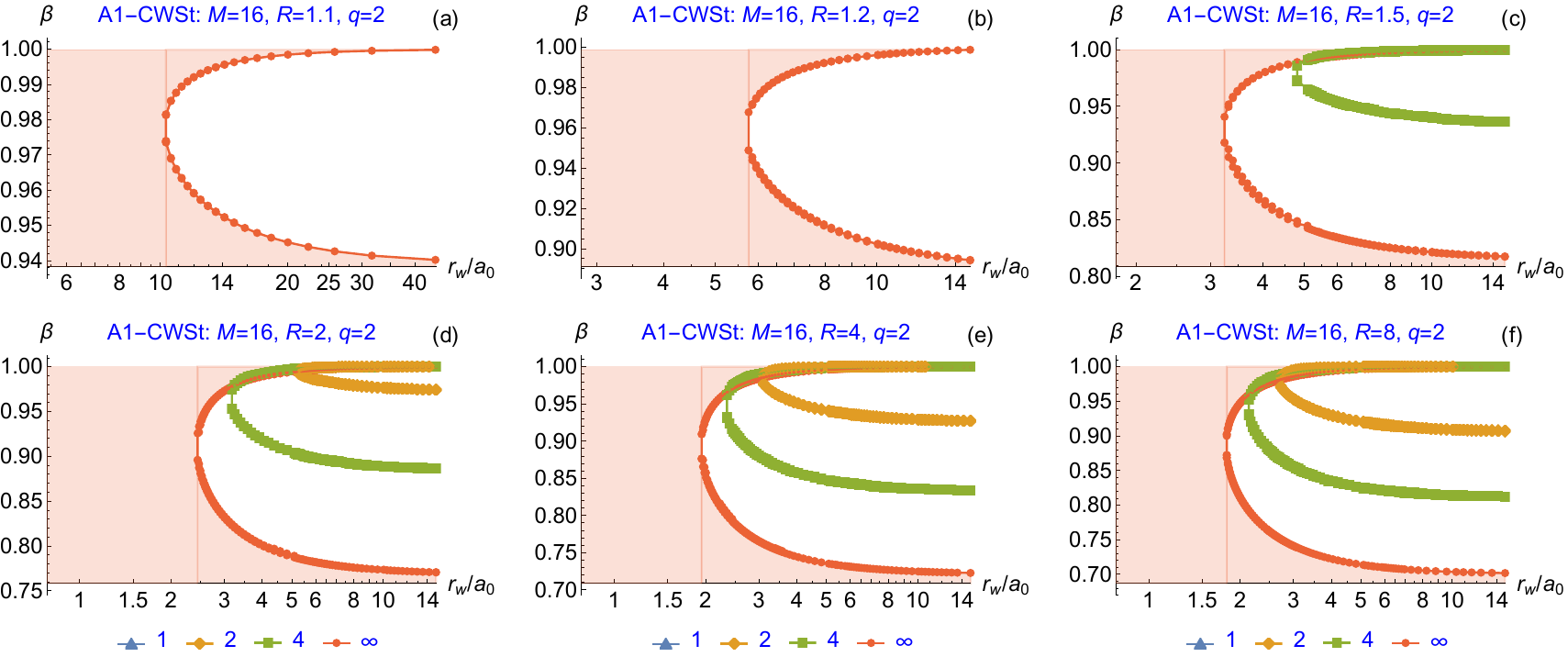}
  \caption{
    Stability map A1-CWSt for model magnetic field \eqref{4:35} and anisotropic plasma pressure model \eqref{3:08} simulating normal NBI at combined stabilization by straight conducting shell and the end MHD anchors; $q = 2$, $M = 16$, various anisotropy $R \in \{1.1, 1.2, 1.5, 2, 4, 8\}$. Instability zone is located between $\beta_{\text{crit}1}(r_{w}/a_{0})$ (lower curve) and $\beta_{\text{crit}2}(r_{w}/a_{0})$ (upper curve of the same color); stability zone is shaded for a plasma with a sharp boundary ($k = \infty $), for which it has the minimum dimensions.
    Compare with Fig.~11\textsuperscript{\cite{Kotelnikov+2023NF_63_066027, Kotelnikov+2023NF_63_066027}} in Ref.~\cite{Kotelnikov+2023NF_63_066027, Kotelnikov+2023NF_63_066027}.
  }
  \label{fig:GM2-A1-CWSt_beta_vs_rw_q2_M16}.
%
\medskip
\centering
\includegraphics[width=\linewidth]{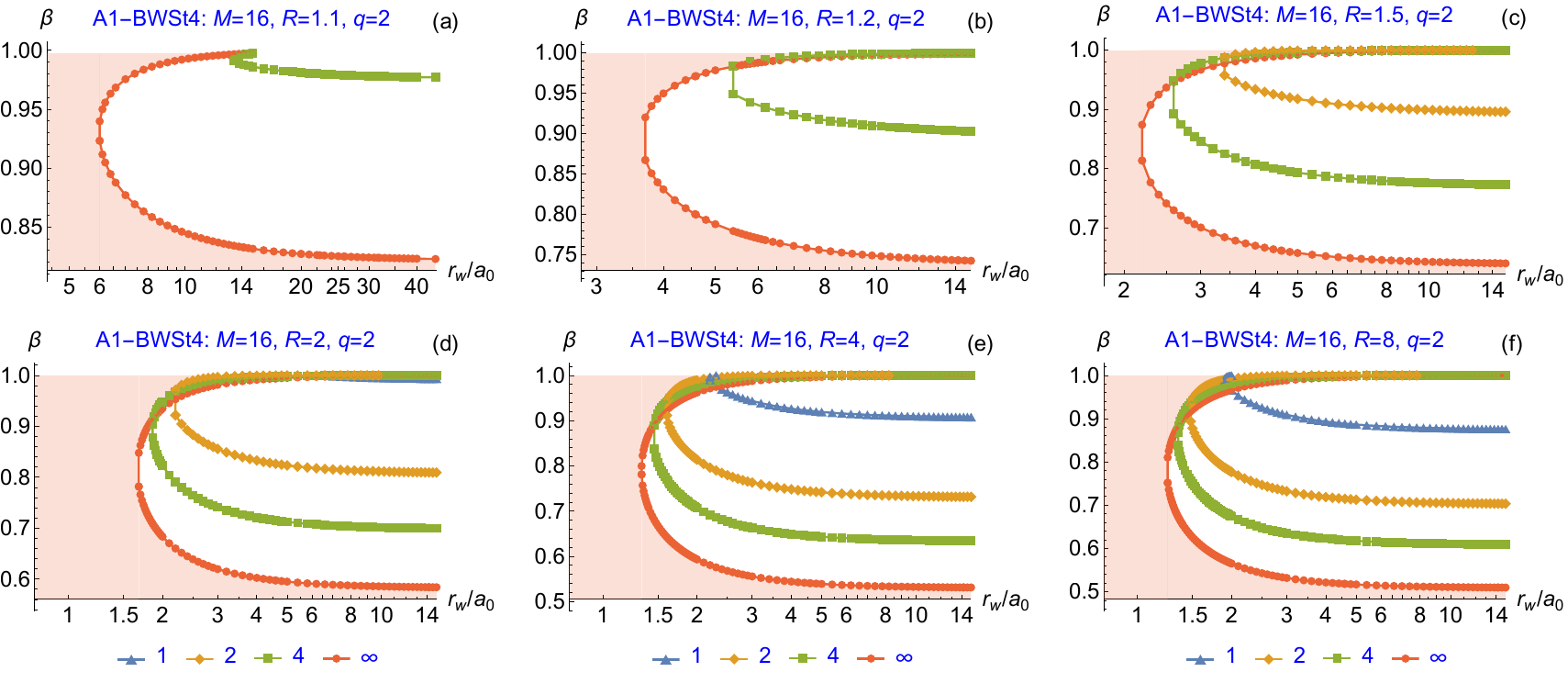}
  \caption{
    Stability map A1-BWSt4 for model magnetic field \eqref{4:35} and anisotropic plasma pressure model \eqref{3:08} simulating normal NBI at combined stabilization by straight conducting shell and the end MHD anchors located beyond mirror throat at  mirror ratio $S=4$; $q = 2$, $M = 16$, various anisotropy $R \in \{1.1, 1.2, 1.5, 2, 4, 8\}$. Instability zone is located between $\beta_{\text{crit}1}(r_{w}/a_{0})$ (lower curve) and $\beta_{\text{crit}2}(r_{w}/a_{0})$ (upper curve of the same color); stability zone is shaded for a plasma with a sharp boundary ($k = \infty $), for which it has the minimum dimensions.
    Correspondence of the colors and markers of the curves to the index $k$ is shown under the bottom row of the graphs.
    Compare with Fig.~\ref{fig:GM2-A1-CWSt_beta_vs_rw_q2_M16}.
  }
  \label{fig:GM2-A1-BWSt_beta_vs_rw_q2_M16}.
\end{figure*}


Following same scheme as in Refs.~\cite{Kotelnikov+2022NF_62_096025, Kotelnikov+2023NF_63_066027, Kotelnikov+2023NF_63_066027}, we performed a series of calculations for the vacuum magnetic field \eqref{4:35} with three values of the index $q\in\{2,4,8\}$ and a mirror ratio from the limited set $M \in \{24, 16, 8, 4\}$. Parameter of anisotropy $R\leq M$ was taken from the list $R\in\{1.1, 1.2, 1.5, 2, 4, 8, 16, 24\}$.


A series of graphs in Fig.~\ref{fig:GM2-A1-CWSt_beta_vs_rw_M=R} illustrates the results of calculations for the case $R=M$, when the plasma anisotropy is minimal and the instability zone has its maximum dimensions. Such a zone in this and subsequent figures lies between the lower and upper branches of each curve, the color of which is individual for each pressure profile with a given index $k\in\{1,2,4,\infty\}$. The lower and upper branches of such a curve are $\beta_{\text{crit}1}$ and $\beta_{\text{crit}2}$, respectively. If only one branch is shown, as in the case of the blue curve for the $k=1$ profile in Fig.~\ref{fig:GM2-A1-CWSt_beta_vs_rw_M=R}(a), we interpret it as $\beta_{\text{crit} 1}$. If there is no curve of a certain color on the graph at all, then we consider that the corresponding pressure profile is stable in the entire range $0<\beta<1$ of beta and in the entire range of ratios $r_{w}/a_{0 }$ on this graph. Recall that $a_{0}$ is the radius of the plasma column in the middle plane of the device, where $a(z)$ is maximum.
%
%


In the region of relatively small ratios $r_{w}/a_{0}$ to the left of the merging point of the $\beta_{\text{crit}1}$ and $\beta_{\text{crit}2}$ branches for the steepest radial profile with index $k=\infty $, on each of the graphs all radial profiles are stable over the entire interval $0<\beta<1$. It is easy to see that smooth pressure profiles ($k=1$, $k=2$) are more stable than steep profiles ($k=4$, $k=\infty$). The same trend also takes place for small-scale ballooning disturbances when end-cap MHD stabilizers are used \cite{Kotelnikov+2021PST_24_015102}.

The graphs in Fig.~\ref{fig:GM2-A1-CWSt_beta_vs_rw_M=R} should be compared with the graphs in Fig.~9\textsuperscript{\cite{Kotelnikov+2023NF_63_066027, Kotelnikov+2023NF_63_066027}} in Ref.~\cite{Kotelnikov+2023NF_63_066027, Kotelnikov+2023NF_63_066027}. As expected, in the straight chamber (that is, in Fig.~\ref{fig:GM2-A1-CWSt_beta_vs_rw_M=R}), the zone of instability is definitely smaller. Most noticeably, the coordinate $r_{w}/a_{0}$ of the left edge of the graphs in the first figure is noticeably smaller than the coordinate $r_{w}/a$ of this edge in the second figure (that is, in the proportional chamber). In other words, in a straight chamber, the vacuum gap (measured in the median plane) must be smaller in order to ensure the stability of a plasma with any radial profile for any value of $\beta $. It is also easy to see that for comparable values of $r_{w}/a_{0}$ in the first figure and $r_{w}/a$ in the second figure, the lower stability zone is $0<\beta <\beta_{\text{crit} 1}$ in the straight chamber is definitely narrower.



The graphs in Fig.~\ref{fig:GM2-A1-CWSt_beta_vs_rw_q2_M16} illustrate the evolution of the margins between the stability and instability zones as the degree of plasma anisotropy changes. This figure should be compared with figure 11\textsuperscript{\cite{Kotelnikov+2023NF_63_066027, Kotelnikov+2023NF_63_066027}} in Ref.~\cite{Kotelnikov+2023NF_63_066027, Kotelnikov+2023NF_63_066027}. The differences between these patterns are minimal at maximum anisotropy (which corresponds to plots (a), $R=1.1$) and maximum at minimal anisotropy (plots (f), $R=8$).


End perfectly conducting plates simulating the effect of the end MHD stabilizers can be located not only in the throat of magnetic mirror. For example, in the GDT facility, has ring limiters (see, e.g., \cite{Bagryansky+2011FST_59_31}). They are located between the magnetic mirror and the stop point of fast ions. Such a limiter in the context of rigid mode stability can be imitated by a conducting end plate installed in a plane with local mirror ratio $S$ such that $R<S<M$. Generally speaking, end plate can be placed both in front of the magnetic plug and behind it. In our internal classification, a configuration with a ring limiter in the region $0<z<1$ is designated `RW' (from the words `Ring Wall') with the addition of a numerical value $S$, while configurations with a conducting end plate behind the magnetic mirror, that is, in the region $ z>1$, has the designation `BW' (from the words `Blind Wall') also with the addition of the number $S$.


Below we assume that the end conducting plates are moved from the magnetic mirror to the behind-mirror region, namely, to the plate where the current mirror ratio is $S=4$. It means that the boundary condition $\phi=0$ should be set not in the plane $z=\pm1$, as Eq.~\eqref{2:11} prescribes, but in the plane $z=z_{e}$, where $B_{v}(z_{e})=S/R$, as prescribed by Eq.~\eqref{2:11S}. A similar transfer of the boundary condition was previously used by \cite{Kesner1985NF_25_275} to imitate an MHD stabilizer that creates a moderate margin of stability. The results of such a calculation are shown in Fig.~\ref{fig:GM2-A1-BWSt_beta_vs_rw_q2_M16}. Comparing Figs.~\ref{fig:GM2-A1-CWSt_beta_vs_rw_q2_M16} and~\ref{fig:GM2-A1-BWSt_beta_vs_rw_q2_M16}, it is easy to see that the second of them has definitely larger instability zones.



In contrast to the wall stabilization shown in Figs.~\ref{fig:GM2-A1-beta_vs_R-L500}, \ref{fig:GM2-A1-LWSt_beta_vs_rw_setAll_M4qs}, \ref{fig:GM2-A1-LWSt_beta_vs_rw_setAll_M8qs} and \ref{fig:GM2-A1-LWSt_beta_vs_rw_setAll_M16qs} in section \ref{s5}, Fig.~\ref{fig:GM2-A1-CWSt_beta_vs_rw_M=R} demonstrates a strong dependence of the instability zone size on the magnetic field profile characterized by the index $q$. As can be seen from Fig.~\ref{fig:GM2-A1-CWSt_beta_vs_rw_M=R}, the width of the instability zone $\beta_{\text{crit}1} < \beta < \beta_{\text{crit}2}$ are maximum for the most smooth magnetic field profile with index $q=2$. It is minimal at $q=8$.

\begin{figure*}
\includegraphics[width=\linewidth]{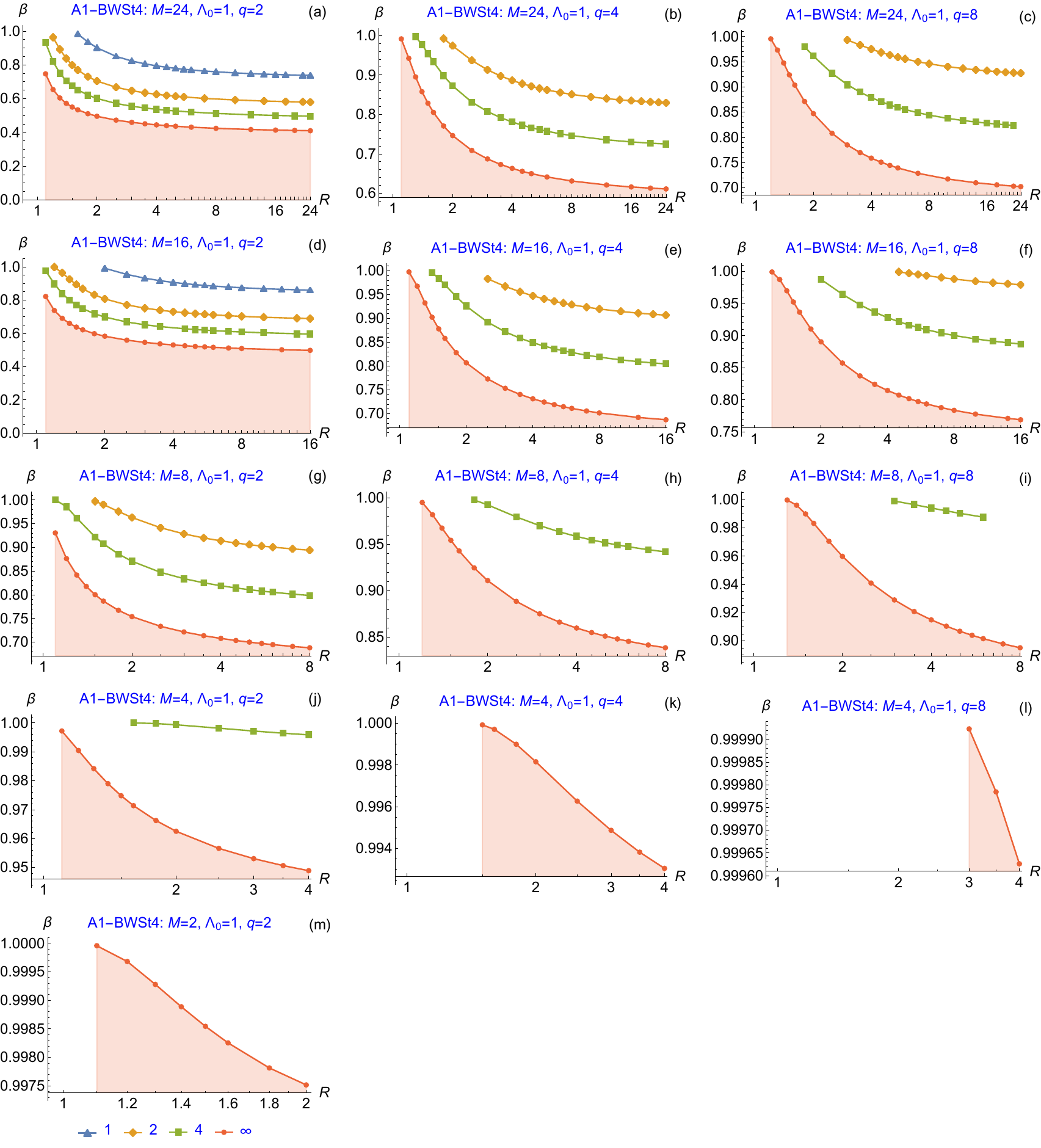}
  \caption{
    Boundary of lower stability zone $\beta_{\text{crit}1}$ for model magnetic field \eqref{4:35} and anisotropic plasma model \eqref{3:08} simulating normal injection of neutral beams, with stabilization by end conducting plates installed outside the central cell at mirror ratio $S=4$; no lateral conducting wall is assumed. Stability zone for the radial profile with index $k$ is located below the corresponding curve. The absence of a curve with particular color means that the corresponding profile is stable over the entire interval $0<\beta<1$.
    Correspondence of the colors and markers of the curves to the index $k$ is shown under the bottom row of the graphs.
    Compare with Fig.~10\textsuperscript{\cite{Kotelnikov+2023NF_63_066027, Kotelnikov+2023NF_63_066027}} in Ref.~\cite{Kotelnikov+2023NF_63_066027, Kotelnikov+2023NF_63_066027}.
  }
  \label{fig:GM2-A1-BWSt4_beta_vs_R}
\end{figure*}
%
%
The shape of the chamber obviously doesn't matter in the limit $r_{w}/a_{0}\to\infty$, since this limit means that the conducting lateral wall is actually removed. We would reproduce Fig.~12\textsuperscript{\cite{Kotelnikov+2023NF_63_066027, Kotelnikov+2023NF_63_066027}} from Ref.~\cite{Kotelnikov+2023NF_63_066027, Kotelnikov+2023NF_63_066027} if we repeat the calculation in this limit for the straight chamber. That figure shows graphs of $\beta_{\text{crit}1}$ versus $R$ for different values of the mirror ratio $M$ and index $q$ in the limit $r_{w}/a=\infty $.



When discussing Fig.~12\textsuperscript{\cite{Kotelnikov+2023NF_63_066027, Kotelnikov+2023NF_63_066027}} in that article, it was said that at a large distance of the lateral conducting wall from the plasma, the upper stability zone becomes extremely thin (or even disappears for radially smooth profiles), and it is located in the immediate vicinity of the boundary $\beta=1$, where the paraxial approximation doesn't work. In this case, the lower zone remains sufficiently wide, which is in full agreement with the results of calculating the threshold of small-scale ballooning instability \cite{BushkovaMirnov1986VANT_2_19e, RyutovStupakov1981IAEA_1_119, Kotelnikov+2021PST_24_015102}. According to those calculations, an analogue of $\beta _{\text{crit}1}$ for such instability even without taking into account the effects of FLR, is estimated to be of the order of $0.6\div0.7$.


Unexpectedly, the analysis of that figure showed that a large-scale ballooning perturbation with an azimuth number $m=1$ in a plasma with a fairly smooth radial profile can be stable over the entire allowable range $0<\beta<1$ of the beta parameter even in the absence of a lateral conducting wall. This requires that the degree of anisotropy be sufficiently high, and the mirror ratio, on the contrary, not too large, and magnetic field profiles with narrow and steep mirrors are more stable.


In order not to tritely repeat figure 12\textsuperscript{\cite{Kotelnikov+2023NF_63_066027}}, we once again assumed that the end conducting plates were rearranged from the throat of magnetic mirror to a locatio with mirror ratio $S=4$. The results are shown in fig.~\ref{fig:GM2-A1-BWSt4_beta_vs_R}. Their comparison with Fig.~12\textsuperscript{\cite{Kotelnikov+2023NF_63_066027}} shows that the decrease in $\beta_{\text{crit}1}$ is very significant and can reach 30. In addition, the range of values of the mirror ratio for which the entire range $0<\beta <1$ is stable for smooth pressure profiles has been significantly reduced. Such a trend was quite expected for a weaker end-face MHD stabilizer.


It may be surprising that the instability zone disappears at $r_{w}/a\to\infty$ as the radial pressure profile is smoothed out. On the one hand, such a statement may seem doubtful. On the other hand, as noted above, even for small-scale perturbations, the value of $\beta_{\text{crit}1}$ is quite large. Intuitively, it seems that with respect to the rigid ballooning mode, the value of $\beta_{\text{crit}1}$ should be even larger, since it is more difficult to deform the entire plasma column than a thin magnetic tube.

\begin{figure*}
  \centering
\includegraphics[width=\linewidth]{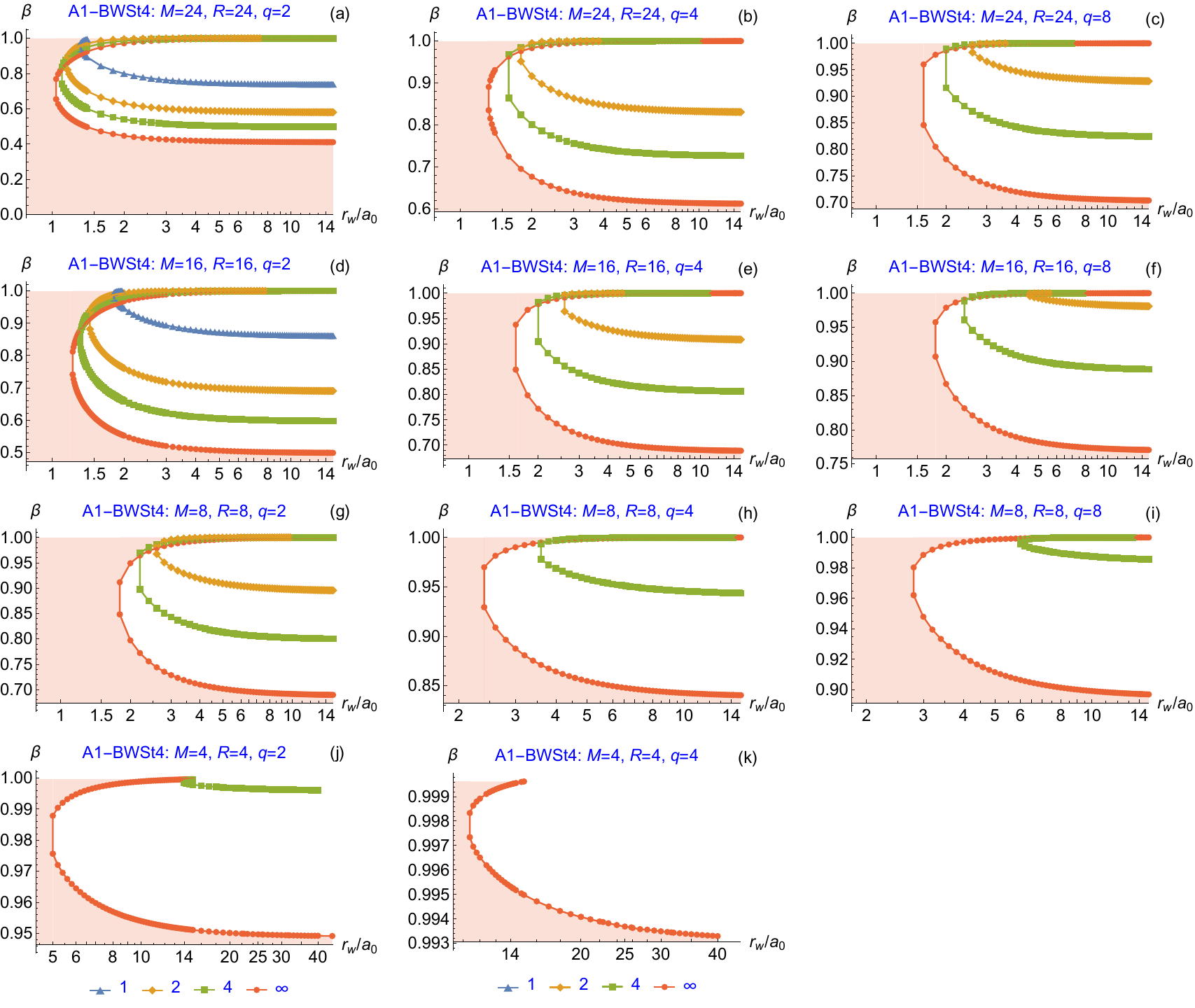}
  \caption{
    Stability map A1-BWSt4 for model magnetic field \eqref{4:35}, straight conducting wall and anisotropic plasma pressure model \eqref{3:08} simulating normal NBI at combined stabilization by weak MHD anchors simulated by conducting end plates installed at mirror ratio $S=4$ behind the magnetic mirror throat;
    $q\in\{2,4,8\}$, different mirror ratios $M\in\{24,16,8,4\}$ and minimum anisotropy $R=M$.
    Instability zone is located between $\beta_{\text{crit}1}(r_{w}/a_{0})$ (lower curve) and $\beta_{\text{crit}2}(r_{w}/a_{0})$ (upper curve of the same color); it is not shaded for a plasma with a sharp boundary ($k=\infty $), for which it has the maximum dimensions.
    Correspondence of the colors and markers of the curves to the index $k$ is shown under the bottom row of the graphs.
    Compare with Fig.~\ref{fig:GM2-A1-CWSt_beta_vs_rw_M=R}.
  }
  \label{fig:GM2-A1-BWSt4_beta_vs_rw_R=M}
\end{figure*}

\begin{figure*}
  \centering
\includegraphics[width=\linewidth]{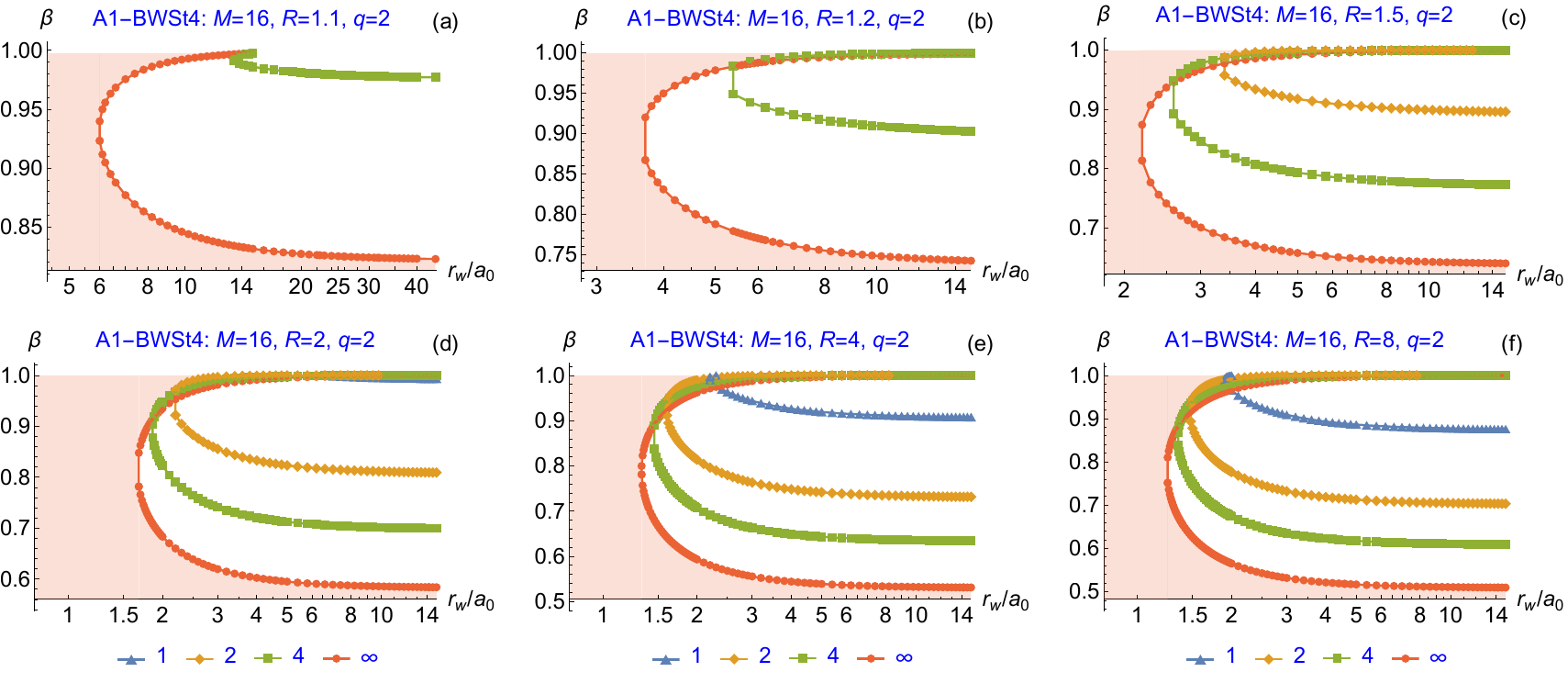}
  \caption{
    %
    Stability map A1-BWSt4 for model magnetic field \eqref{4:35}, straight conducting wall and anisotropic plasma pressure model \eqref{3:08} simulating normal NBI at combined stabilization by by weak MHD anchors simulated by conducting end plates installed at mirror ratio $S=4$ behind the magnetic mirror throat;
    $q\in\{2\}$, $M\in\{16\}$, various anisotropy $R\in\{1.1, 1.2, 1.5, 2,4,8\}$.
    Instability zone is located between $\beta_{\text{crit}1}(r_{w}/a_{0})$ (lower curve) and $\beta_{\text{crit}2}(r_{w}/a_{0})$ (upper curve of the same color); it is not shaded for a plasma with a sharp boundary ($k=\infty $), for which it has the maximum dimensions.
    Correspondence of the colors and markers of the curves to the index $k$ is shown under the bottom row of the graphs.
    Compare with Fig.~\ref{fig:GM2-A1-CWSt_beta_vs_rw_q2_M16}.
  }
  \label{fig:GM2-A1-BWSt4_beta_vs_rw_q2_M16}
\end{figure*}


The last two figures \ref{fig:GM2-A1-BWSt4_beta_vs_rw_R=M} and \ref{fig:GM2-A1-BWSt4_beta_vs_rw_q2_M16} are drawn as analogues of Figs.~\ref{fig:GM2-A1-CWSt_beta_vs_rw_M=R} and~\ref{fig:GM2-A1-CWSt_beta_vs_rw_q2_M16} with the difference that, like Fig.~\ref{fig:GM2-A1-BWSt4_beta_vs_R}, they display the simulation result of a weakened MHD stabilizer with $S=4$ . Comparing these pairs of figures, it can again be seen that the weakening of the end MHD stabilizer expands the zone of instability, so that smoother radial pressure profiles become unstable.

%

\medskip

Summing up everything said in this section, we can state the following:
\setlength{\itemindent}{0em}
\setlength{\leftmargini}{1em}
\begin{itemize}
  \item
    %
    Two stability zones are found for moderate values of the ratio $r_{w}/a_{0}$ and a sufficiently large mirror ratio $M$. The lower stability zone $\beta < \beta_{\text{crit}1}$ exists even if the lateral conducting wall is removed, that is, in the case of $r_{w}/a_{0}=\infty $.
  \item
    %
    Other things being equal, the instability zone is maximum for the steepest radial pressure profile ($k=\infty $) and may be completely absent for fairly smooth radial profiles ($k=1$ or $k=2$). On the contrary, as shown in section \ref{s5}, it is for a profile with a sharp boundary that the instability zone had the minimum dimensions.

  \item
    %
    For a fixed ratio $r_{w}/a_{0}$, the stability zones expand and can merge with a decrease in the mirror ratio $M$ and/or a smoothing of the radial pressure profile (with a decrease in $k$).

  \item
    %
    The instability zone decreases and may even disappear as the plasma anisotropy increases (as $R$ decreases). The dimensions of the unstable zone are maximal at the minimum degree of anisotropy $R=M$ admissible for a given ratio of mirrors $M$.

  \item
    %
    If an instability zone exists between two stability zones for some combinations of the parameters $k$, $q$, $M$, and $R$, then it disappears if $r_{w}/a_{0}$ is below some threshold.

%

  \item
    %
    With a not too large mirror ratio, and/or a sufficiently high plasma anisotropy, and/or a sufficiently steep magnetic field, and/or a sufficiently smooth pressure profile, and/or sufficiently close lateral wall, the rigid ballooning mode $m=1$ can be stabilized at any feasible value of beta.

  \item
    %
    Zone of instability noticeably expands as the stability margin created by the end MHD anchors decreases.

\end{itemize}

\medskip
\section{Conclusions}\label{s9}


In this paper, within the framework of an anisotropic pressure model simulating the normal injection of fast neutral beams into a relatively cold target plasma, we study the possibility of stabilizing the $m=1$ rigid flute and ballooning modes in an axially symmetric mirror trap using a perfectly conducting cylindrical wall in the form of a straight cylindrical chamber.  Unlike most previous works, which assumed a not quite realistic plasma pressure profile in the form of a step with a sharp boundary, and the shape of the conducting chamber repeated the shape of the plasma column on an enlarged scale, we considered a set of diffuse pressure profiles with different degrees of sharpness of the plasma edge, as well as several variants of the axial profile of the vacuum magnetic field with different mirror ratios and different slopes of the magnetic mirrors.

As expected, the conducting wall in the form of a straight cylinder has a weaker stabilizing effect on ballooning perturbations compared to a proportional chamber, but the attenuation does not seem to be critically large. It can be assumed that under the conditions of a real experiment, the shape of the conducting chamber will be something in between the proportional and straight chambers. In this sense, these two chambers are two antipodes, and the results of calculations for these chambers determine the boundaries of the interval of the expected value $\beta_{\text{crit}}$ for a real experiment.


Admitting as more than encouraging the conclusion about the weak influence of the shape of the conducting chamber on its stabilizing properties, it is necessary to make a reservation that this conclusion is based on the analysis of only one model of anisotropic pressure, which imitates the plasma that is formed by normal NBI to the minimum of the magnetic field at the right angle to the axis of the trap. It should be expected that with oblique injection, the influence of the shape of the lateral conducting wall will increase markedly.

\begin{acknowledgments}


    This work has been done in the framework of ALIANCE collaboration \cite{Bagryansky+2020NuclFusion_60_036005}. It was supported by Chinese Academy of Sciences President’s International Fellowship Initiative (PIFI) under the Grants No.~2022VMA0007
    and Chinese Academy of Sciences International Partnership Program under the Grant No.~116134KYSB20200001.


\end{acknowledgments}

\section*{ORCID iDs}


\noindent
Igor KOTELNIKOV \href{https://orcid.org/0000-0002-5509-3174}{https://orcid.org/0000-0002-5509-3174}



%

\end{document}